%Paper: hep-th/9410051
%From: sezgin@bose.tamu.edu
%Date: Fri, 07 Oct 1994 12:07:05 CST
%Date (revised): Tue, 8 Nov 1994 20:10:31 -0500

%%%%%%%%%%%%%%%%%%%%%%%%%%%%%%%%%%%%%%%%%%%%%%%%%%%%%%%%%%%%%%%%
%%                                                            %%
%%                                                            %%
%%       Search for Duality Symmetries in p-Branes            %%
%%                                                            %%
%%                     E. Sezgin                              %%
%%                                                            %%
%%                    PLAIN TEX                               %%
%%                                                            %%
%%%%%%%%%%%%%%%%%%%%%%%%%%%%%%%%%%%%%%%%%%%%%%%%%%%%%%%%%%%%%%%%

\magnification=\magstep1
\baselineskip=15pt
\font\fontina=cmr9
\nopagenumbers

\def\p{\partial}
\def\py{\partial y}

\def\tr{{\rm tr}}
\def\det{{\rm det}}

\overfullrule=0pt

\line{\hfil CTP TAMU-45/94}
\line{\hfil hep-th/9410051}
\line{\hfil October 1994}
\vskip 1truecm
\centerline{
{\bf Search for Duality Symmetries in p-Branes}
\footnote{$^*$}{\fontina Talk presented at the
 {\it G\"ursey\ Memorial\ Conference\ I\ on\ Strings\ and\ Symmetries},
Istanbul, Turkey, June 6-10, 1994.}
}
\vskip 1truecm
\centerline{E. Sezgin }
\bigskip
\centerline{\it Center for Theoretical Physics, Texas A\&M University,}
\centerline{\it College Station, Texas 77843-4242, U.S.A.}
\vskip 2.5truecm
\centerline{\bf Abstract}
\smallskip
\midinsert\narrower{ The requirement of an $SL(2)$
duality  symmetry,
mixing the worldvolume field equations with Bianchi
identities, leads to a highly nonlinear equation involving the
transformation parameters and certain worldvolume currents. In
general, this  equation
seems to  admit a solution only for a two parameter subgroup of the
seeked $SL(2)$. These transformations also leave invariant the first
class constraints generating  the worldvolume reparametrizations. In
the special case of $p$--branes in $p+1$ dimensions, the full $SL(2)$
is realized.
}\endinsert
\vfill\eject
\medskip

I met Feza G\"ursey a number of times, but
unfortunately I never had a chance to collaborate with
him. I have always been inspired by his beautiful work and his
kind and gentle personality. I have a great respect and
admiration for him. He is greatly missed and he will always be
remembered.

Feza G\"ursey had a keen sense for beauty of symmetries in physics and
he made profound contributions in search of them. I believe that the
topic of this talk, which is dedicated
to his memory, is very much in the spirit of his
research philosophy that emphasizes symmetries. More specifically, I
will talk about a search for duality symmetries in theories of extended
objects, known as $p$-branes.

It is well known that the ten dimensional heterotic string theory
compactified on a six torus has target space
duality symmetry group $O(6,22;Z)$ which mixes momentum modes with
winding modes [1,2]. From the world-sheet field theory point of view,
this group mixes the world-sheet field equations with Bianchi identities
[2]. It is natural  to search for similar duality symmetries for
higher $p$--branes.

In view of the possibility of a connection beetween the
strongly coupled heterotic string and a weakly coupled fivebrane theory
in ten dimensions [3], the issue of duality symmetry in fivebrane theory
especially seems interesting. The expected target space duality group in
this case is an $SL(2,Z)$ group [4] (For a review, see [5]). Although it
has been shown that an  $SL(2,Z)$
duality group indeed mixes the momentum modes with the winding modes
of the
fivebrane theory [6], so far this symmetry group has not been
understood
as a worldvolume duality group transforming the worldvolume equations of
motion into Bianchi identities. In a recent paper [7], we studied the
problem
from this point of view. We found that, with a reasonable set of
assumptions about the form of the duality transformations, the
existence of an $SL(2)$ symmetry requries a solution to a highly
nonlinear
equation involving the transformation parameters and the worldvolume
currents that transform into each other under duality transformations.
In general, this equation seems to admits only a two parameter solution
which forms a subgroup of the seeked $SL(2)$.  In
the special case of $p$--branes in $p+1$ dimensions, the full $SL(2)$
is realized. In this note, we will
describe
the main result of [7], and we shall furthermore show that the two
parameter duality group also leaves invariant the first class
constraints
which generate the worldvolume reparametrizations. An attempt to find
duality symmetries in $p$--branes has been made before in somewhat
different setting [8]. Further comments on this paper can be found
in [7].

While the case of most interest is the fivebrane, we shall consider all
$p$--branes, without much more effort. The dynamical variables
describing
the $p$--brane are scalar fields $x^\mu(\sigma)$, $y^\alpha(\sigma)$
and a worldvolume metric $\gamma_{ij}(\sigma)$. Here $\sigma^i\
(i=0,...,p)$
are the worldvolume coordinates, $x^\mu$, $\mu=0,...,m-1$ are
coordinates
on $m$ dimensional space $M$ and $y^\alpha$, $\alpha=1,...,p+1$ are
coordinates on a compact $(p+1)$-dimensional manifold $N$.
We shall take the background fields to be the metrics
$g_{\mu\nu}(x)$ and $g_{\alpha\beta}(x)$ on $M$ and $N$ respectively
and an antisymmetric tensor field
$b_{\alpha_1\ldots \alpha_{p+1}}(x)
=\lambda_1(x)\epsilon_{\alpha_1\ldots \alpha_{p+1}}$. In this background,
the usual Polyakov type action for the $p$--brane is
$$
\eqalign{
S =\int d^{p+1}\sigma {\cal L}
= \int d^{p+1}\sigma &
\Bigl[-{1\over2}\sqrt{-\gamma}
\left(\gamma^{ij}\p_i x^\mu\p_j x^\nu g_{\mu\nu}
+\gamma^{ij}\partial_iy^\alpha \partial_jy^\beta g_{\alpha\beta}\right)
+{p-1\over2}\sqrt{-\gamma}\cr
&+{1\over(p+1)!}        \epsilon^{i_1\ldots i_{p+1}}
\p_{i_1}y^{\alpha_1}\cdots\p_{i_{p+1}}y^{\alpha_{p+1}}
\lambda_1\epsilon_{\alpha_1\ldots \alpha_{p+1}}\Bigr]\ .\cr}\eqno(1)
$$
Let us concentrate on the field equation for the internal coordinates
$y^\alpha$. It reads $ \partial_i P^i{}_\alpha=0$, where
$$
\eqalignno{
P^i{}_\alpha=&{\partial{\cal L}\over\partial\partial_i y^\alpha}=
-\sqrt{-\gamma}\gamma^{ij}\partial_j y^\beta g_{\beta\alpha}
+\lambda_1 J^i{}_\alpha\ , &(2)\cr
J^i{}_\alpha=&{1\over p!}\epsilon^{ij_1\ldots j_p}
\p_{j_1}y^{\beta_1}\cdots\p_{j_p}y^{\beta_p}
\epsilon_{\alpha\beta_1\ldots\beta_p}\ .
&(3)\cr}
$$

In searching for a duality symmetry in $p$--brane theories, we also
need to know how the induced metric on the worldvolume transforms.  We
know from ten dimensional supergravity compactified on a six--torus that
under $SL(2)$ the metrics $g_{\alpha\beta}$ and
 $g_{\mu\nu}$ rescale. Therefore let us define
$$
g_{\mu\nu}=\lambda_2^K \bar g_{\mu\nu}\ , \qquad\qquad
g_{\alpha\beta}=\lambda_2^L \bar g_{\alpha\beta}\       , \eqno(4)
$$
where $\bar g_{\alpha\beta}$ and $\bar g_{\mu\nu}$ are assumed to be
inert under $SL(2)$,
and ${\rm det}\ \bar g_{\alpha\beta}=1$. Thus
$\lambda_2(x)=\left({\rm det}
g_{\alpha\beta}\right)^{1/(p+1)L}$.
In the case $p=5$ it is known from the $SL(2)$ duality symmetry of the
effective field theory limit that $K=-1$ and $L=1/3$ [5].

The equation of motion for the worldvolume metric $\gamma$ gives
$$
\gamma_{ij}=\lambda_2^K\p_i x^\mu\p_j x^\nu \bar g_{\mu\nu}+
\lambda_2^L
\p_i y^\alpha\p_j y^\beta \bar g_{\alpha\beta}\ .\eqno(5)
$$
Thus, the duality transformation rule for the worldvolume metric
follows from the transformation rules assigned to quantitities
occuring in this equation.

 We now
look for transformation rules that mix the field equation  $ \partial_i
P^i{}_\alpha=0$, with the Bianchi identity $\partial_i J^i{}_\alpha=0$.
The most natural way to do this is to consider the transformations of
the currents $P^i{}_\alpha$ and $J^i{}_\alpha$ into each other as
$$
\eqalignno{
\delta P^i{}_\alpha=&\,a P^i{}_\alpha+b J^i{}_\alpha \ , &(6a) \cr
\delta J^i{}_\alpha=&\,c P^i{}_\alpha+d J^i{}_\alpha\ ,  &(6b)\cr}
$$
with  $a,b,c,d$ being constants. It is important to take the
parameters to be constant, so that these transformations indeed map the
field equations and Bianchi identities into a combination of each other.
The key point in establishing the duality symmetry is to show that (2)
is invariant under the transformations (6), combined with appropriate
transformation rules for the background fields $\lambda_1$ and
$\lambda_2$. In this regard, we are following the approach of Gaillard
and Zumino [9].

Since all relevant quantities have two indices, it is convenient
to use matrix notation. We define matrices $P$ and $J$
with components $P^i{}_\alpha$ and $J^i{}_\alpha$, matrices
$\py$ with components $(\py)^\alpha{}_i=\p_i y^\alpha$ and
$\bar g^{(p+1)}$ with components $\bar g_{\alpha\beta}$.
{}From the definition (3) we find that $ \p y=J^{-1}(\det J)^{1/p}$.
This equation allows us to calculate the variation of $\p y$ under the
duality transformations.

The central result of [7] is that, taking into account the variation of
the worlvolume metric $\gamma$, and allowing any transformation rules for
the scalar fields $\lambda_1,\lambda_2$, the invariance of (2)
under the duality transformations (6) implies requires that the
following highly nonlinear equation be satisfied:
$$
\eqalign{ &c X^2\ +\
\left[a-2c\lambda_1 -{1\over p}(d+c\tr X)-\left({p-1\over2}K+L \right)
\lambda_2^{-1}\delta\lambda_2 \right]X
\cr
&+b-{p-1\over p}d\lambda_1+{1\over p}c\lambda_1\tr X
+\left(
{p-1\over2}K+L \right)\lambda_1\lambda_2^{-1}\delta\lambda_2
-\delta\lambda_1=
\cr
=&
\lambda_2^L\gamma^{-1}V\Biggl\{
2c X^2-\left[{2\over p}(d+c\tr X)+2c\lambda_1
+\left(L-K\right)\lambda_2^{-1}\delta\lambda_2 \right] X
\cr
&+{1\over p}c(\tr X)^2-c \tr(X^2)
+\left({1\over p}c\lambda_1+{1\over p}d
+{1\over 2} (L-K) \lambda_2^{-1}\delta\lambda_2\right)\tr X \cr
&\qquad\qquad\qquad\qquad
-{p-1\over p}d\lambda_1
-{1\over 2} (p-1)(L-K)\lambda_1\lambda_2^{-1}\delta\lambda_2
\Biggr\}\ , \cr}\eqno(7)
$$
where
$$
X=P\cdot J^{-1}=\lambda_1+\lambda_2^L{\gamma^{-1}V \over
\sqrt{-\det(\gamma^{-1}V)}}\ ,\quad\quad
V=(\partial y)^T\bar g^{(p+1)}\partial y \ . \eqno(8)
$$

Since $\gamma^{-1}V$ can be reexpressed in terms of $X$ via (8),
this is an infinite polynomial equation in the $(p+1)\times(p+1)$
matrix $X$. One is free to determine $\delta\lambda_1$ and
$\delta\lambda_2$ as functions of $a,b,c,d,\lambda_1,\lambda_2$
to satisfy this equation, and also if necessary to put
restrictions on the transformation parameters $a,b,c,d$.
It is important to realize that these transformations have to
be the same for all $X$. In order to prove that this equation
has no solution it would therefore be sufficient to find two
particular matrices $X$ which   give incompatible values
for the variations $\delta\lambda_1$ and $\delta\lambda_2$.

In [7], we implemented this idea by  choosing a particular
matrix $X$ and expanding equation (7) around it.
For our background we choose the fields $x^\mu$ and $y^\alpha$
such that $\gamma^{-1}V=\lambda_2^{-L}\eta$, where $\eta$ is the
Minkowski metric, and then write
$ \gamma^{-1}V=\lambda_2^{-L}(\eta+Y) $. Expanding (7) in powers of
$Y$, we then find that the solution of (7) is given by
$$
\eqalign{
c=&\, 0\ ,\cr
d=&\, p{K-L\over pK+L}a\ , \cr
\delta\lambda_1=& b+{(p+1)L\over pK+L}a\lambda_1\ , \cr
\delta\lambda_2=& { 2\over pK+L}a\lambda_2 \ . \cr}\eqno(9)
$$
In fact, one can check this is the solution of the full equation (7).
Furthermore, one can show that the $x^\mu$-equation of motion is also
invariant under these transformations. Therefore, we have a two
parameter group of duality transformations of the $p$--brane. It is easy
to check that the transformation rules (6) and (9) yield the same
commutator algebra. Denoting the transformations by $a$ and $b$, the
only nonvanishing commutator is $[a,b]=b$.
One can have $d=-a$ by choosing $K={p-1\over 2p}L$.
In this way the two parameter group appears to be a subgroup of the
expected group $SL(2)$. Notice that,
from a solution with magnetic charge, one can obtain a solution
with magnetic and electric charge.
However, the two parameter group does not contain the important
$R\rightarrow 1/R$ transformations.

In the special case of a $p$--brane propagating in $p+1$ dimensions,
it is straightforward to show that the full $SL(2)$ duality
symmetry group is realized.

Finally, let us consider the action of the two parameter duality group
described above on the reparametrization generating constraints.
These constraints are [10]
$$
\eqalignno{
H=&g^{\mu\nu}p_\mu
p_\nu+g^{\alpha\beta}(p_\alpha-\lambda_1 j_\alpha)(p_\beta-\lambda_1
j_\beta)  \cr
&\ \ \ \ \ \ \ \ \ \ \ \ \ +{\rm det}\ (\p_a x^\mu \p_b x^\nu g_{\mu\nu}
+ \p_a y^\alpha \p_b y^\beta g_{\alpha\beta} )\ , &(10)\cr
H_a=&\p_a x^\mu p_\mu+\p_a y^\alpha p_\alpha\ ,  &(11) \cr}
$$
where the index
$a=1,...,p$ labels the spatial directions, $p_\mu$ and $p_\alpha$ are
the momentum variables associated with $x^\mu$ and $y^\alpha$,
respectively, and $ j_\alpha=J^0{}_\alpha$, which can be read off from
$J^i{}_\alpha$ given in (3).

 Consistent with the definition (10) and the
transformations rules (6) and (9) one finds
$\delta \p_a y^\alpha = {d\over p} \p_a y^\alpha. $
Using this variation, and taking $x^\mu$, and therefore $\p_a
x^\mu$ to be inert, we find that under the duality transformations (6)
and (9) the constraint $H_a$ is preserved: $\delta H_a ={(p+1)K\over
pK+L}a H_a$, provided that we assign the transformation rule
$
\delta p_\mu={(p+1)K\over pK+L}a p_\mu. $
Finally, one finds that the constraint $H$ is also preserved under the
duality transformation: $\delta H= {pK\over pK+L}a H$.

In the special case of $p$--branes in $(p+1)$--dimensions, the
Hamiltonian simplifies drastically to
$$
H=g^{\alpha\beta}(p_\alpha-\lambda_1
j_\alpha)(p_\beta-\lambda_1 j_\beta)+
\lambda_2^{(p+1)L}g^{\alpha\beta}j_\alpha j_\beta\ . \eqno(12)
$$
One can show that the Hamiltonian constraint and the space
reparametrization constraint
$H_a=\p_a x^\mu p_\mu+\p_a y^\alpha p_\alpha$ are preserved under the
full $SL(2)$ duality transformations.
\bigskip
\centerline{\bf Acknowledments}
\bigskip
I would like
to thank the organizers of the {\it G\"ursey\ Memorial\ Conference\ I\
on\ Strings\ and\ Symmetries} for their kind invitation. I also thank
Roberto Percacci, with whom we collaborated on the material summarized
here (see ref. [7]). This work has been supported in part by the U.S.\
National Science Foundation, under grant PHY-9106593.
\vfill\eject
\medskip
\centerline{\bf References}
\vskip 0.75truecm

\item{[1]} K.S. Narain, M.H. Sarmadi and E. Witten,
Nucl. Phys. {\bf B279}
                (1987) 367;
\item{} A. Shapere and F. Wilczek, Nucl. Phys. {\bf B320} (1989) 669;
\item{} A. Giveon, E. Rabinovici and G. Veneziano, Nucl. Phys. {\bf B322}
               (1989) 167;
\item{} A. Giveon, N. Malkin and E. Rabinovici, Phys. Lett. {\bf 220}
       (1989) 551;
\item{} W. Lerche, D. L\"ust and N.P. Warner, Phys. Lett. {\bf B231}
        (1989) 417.
\item{[2]} M.J. Duff, Phys. Lett. {\bf B173} (1986) 289;
\item{} S. Cecotti, S. Ferrara and L. Girardello, Nucl. Phys. {\bf B308} (1988)
               436;
\item{} J. Molera and B. Ovrut, Phys. Rev. {\bf D40} (1989) 1146;
\item{} M.J. Duff, Nucl. Phys. {\bf B335} (1990) 610;
\item{} J. Maharana and J.H. Schwarz, Nucl. Phys. {\bf B390} (1993) 3.
\item{[3]} M.J. Duff, Class. Quant. Grav. {\bf 5} (1988) 189;
\item{} A. Strominger, Nucl. Phys. {\bf B343} (1990) 167;
\item{} M.J. Duff and J.X. Lu, Nucl. Phys. {\bf B354} (1991) 141;
        Phys. Rev.      Lett. {\bf 66} (1991) 1402; Class. Quant. Grav.
       {\bf 9} (1991) 1;
\item{} C.G. Callan, J.A. Harvey and A. Strominger, Nucl. Phys.
      {\bf B359}(1991) 611; Nucl. Phys. {\bf B367} (1991) 60;
\item{} M.J. Duff, R. Khuri and J.X. Lu, Nucl. Phys. {\bf B377} (1992)
        281;
\item{} J. Dixon, M.J. Duff and J. Plefka, Phys. rev. Lett.
       {\bf  69} (1992) 3009.
\item{[4]} J.H. Schwarz and A. Sen, Phys. Lett. {\bf B312} (1993) 105;
\item{} J.H. Schwarz and A. Sen, Phys. Lett. {\bf B312} (1993) 105;
\item{} P. Binetr\'uy, Phys. Lett. {\bf B315} (1993) 80.
\item{[5]}      A. Sen, preprint, TIFR/TH/94-03 (hep-th/9402002).
\item{[6]} A. Sen, Nucl. Phys. {\bf B388} (1992) 457;
                     Phys. Lett. {\bf B303} (1993) 22;
                Mod. Phys. Lett. {\bf A8} (1993) 2023;
 \item{} J.H. Schwarz and A. Sen, Phys. Lett. {\bf 312} (1993) 105;
\item{} T. Ortin, Phys. Rev. {\bf D47} (1993) 3136;
\item{} M.J. Duff and R. Khuri, preprint, CTP-TAMU-17/93.
\item{[7]} R. Percacci and E. Sezgin, preprint, CTP TAMU-15/94, SISSA
          44/94/EP (hep-th/9407021).
\item{[8]} M.J. Duff and J.X. Lu, Nucl. Phys. {\bf B347} (1990) 394.
\item{[9]} M.K. Gaillard and B. Zumino, Nucl. Phys. {\bf B193} (1981)
          221.
\item{[10]} E. Bergshoeff, R. Percacci, E. Sezgin, K.S. Stelle and
           P.K. Townsend, Nucl. Phys. {\bf B398} (1993) 343.
\end